# WebProtégé: A Cloud-Based Ontology Editor

Matthew Horridge
Stanford University
matthew.horridge@stanford.edu

Rafael S. Gonçalves
Stanford University
rafael.goncalves@stanford.edu

Csongor I. Nyulas
Stanford University
nyulas@stanford.edu

Tania Tudorache
Stanford University
tudorache@stanford.edu

Mark A. Musen
Stanford University
musen@stanford.edu

## ABSTRACT

We present WebProtégé, a tool to develop ontologies represented in the Web Ontology Language (OWL). WebProtégé is a cloud-based application that allows users to collaboratively edit OWL ontologies, and it is available for use at https://webprotege.stanford.edu. WebProtégé currently hosts more than 68,000 OWL ontology projects and has over 50,000 user accounts. In this paper, we detail the main new features of the latest version of WebProtégé.

## 1 INTRODUCTION

The Protégé team at the Stanford Center for Biomedical Informatics Research has been developing software for formally representing knowledge in the form of ontologies since the early 90s [3]. With the growing adoption of logic-based knowledge representation languages that have precisely-defined semantics—such as the World Wide Web Consortium (W3C) standard Web Ontology Language (OWL) [1]—the WebProtégé [2, 5] software has enjoyed increasing traction and demand from its growing user base. In particular, the recent popularity of so-called "knowledge graphs" has brought a new wave of users to the WebProtégé user community, as well as students to our Protégé courses. Ontologies can be used as components of knowledge graphs, to equip them with precise semantics and to enable logical reasoning. We co-developed multiple ontologies that are used in enterprise knowledge graphs, and some of the features in WebProtégé we present here were driven by those experiences. With this paper, we hope to draw feedback from the knowledge graph community so that we can steer our development efforts to better suit the needs of this community.

The WebProtégé cloud-based tool supports simultaneous collaborative editing of OWL ontologies for geographically distributed users. By design, WebProtégé offers a simple user interface streamlined for common ontology engineering patterns and tasks. WebProtégé is available for general purpose use at https://webprotege.stanford.edu, where it hosts over **68,000 OWL ontology projects** and over **50,000 user accounts**. The software is open-source and all source code is available on GitHub.[1] In its latest version, WebProtégé 4.0 builds on the success of previous versions but features a new modern interface, and new features for collaboration, search, and quality assurance.

## 2 WEBPROTÉGÉ CLOUD ONTOLOGY EDITOR

WebProtégé is essentially a "Google Docs" for editing ontologies. Users access "projects" that are collections of OWL ontologies augmented with a change history, and issues/comments. WebProtégé provides a default user interface that supports lightweight ontology editing, for ontologies that more or less fit within the OWL2EL profile.[2] However, it can also be configured for editing ontologies that require full-blown OWL 2 axioms[3] and class expressions.

### 2.1 Ontology Editing

Figure 1 shows the main editing interface of WebProtégé, which provides a tabbed look and feel. The classes tab comprises four resizable views: Class Hierarchy, Class Description, Comments and Project Feed. Users can add, remove and customize the layout of views on a tab, and the state of the user interface is automatically persisted for each user in the context of a project.

In contrast to previous versions, WebProtégé 4.0 uses proper URL routing, meaning that entities, such as classes, properties and individuals, and the tabs (or pages) that display them, are bookmarkable links that can be shared with people.

WebProtégé allows ontology edits to be seen by all users as they take place. It is possible to set up a project so that users may only view or comment on a project for cases where there is a core group of editors and a larger group of commenters/reviewers.

Users of WebProtégé can easily link to entities in popular knowledge bases such as **Schema.org**, **Wikidata**, and **DBpedia**. For example, to link out to the entity Dataset from Schema.org, a user types in schema:Dataset when creating a new class and WebProtégé renders the entity to its IRI https://schema.org/Dataset, and similarly for Wikidata and DBpedia entities using prefix names wikidata and dbpedia, respectively.

[1] https://github.com/protegeproject/webprotege
[2] https://www.w3.org/TR/owl2-profiles/#OWL_2_EL
[3] https://www.w3.org/TR/owl2-syntax/

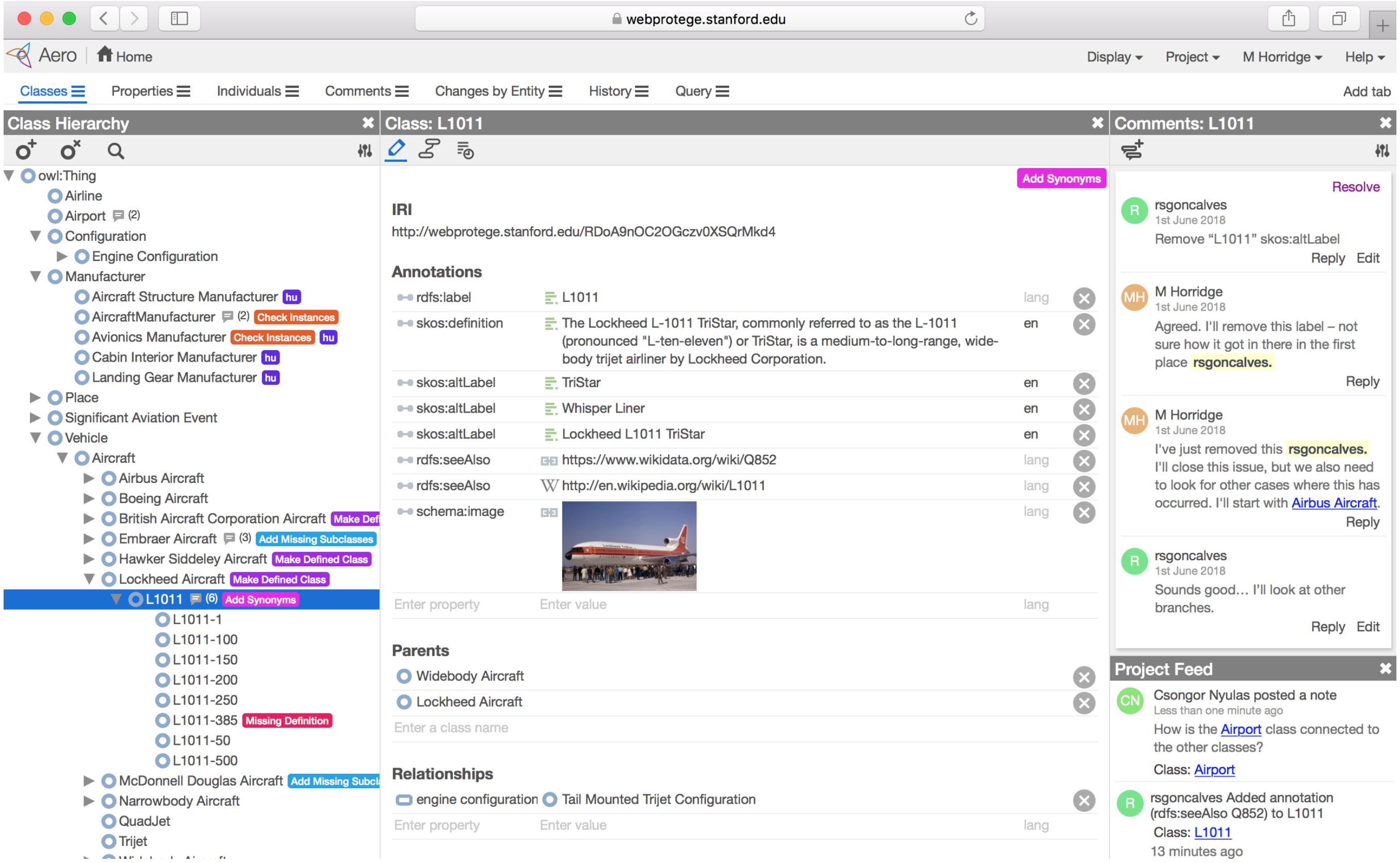

**Figure 1: Class editing in WebProtégé. This figure demonstrates the default user interface and shows the use of tags, comments/issues and the project feed.**

## 2.2 Collaborative Features

Besides collaborative ontology editing functionality, WebProtégé also allows users to comment on a project and create issues in the form of *threaded comments*. Comments sit outside of an ontology but pertain to specific entities in an ontology. They are formatted in Markdown syntax and can contain links to any entities in the ontology as well as GitHub style @user mentions.

Discussion threads of comments for a given entity can be seen on the right-hand side of Figure 1 in the Comments view. The number of threads is also displayed next to entity names in the various entity hierarchies (left hand side of Figure 1). Threads can be created and closed as issues are opened and then dealt with by ontology editors. Figure 2 shows a global view of the Comments tab that is available in the default interface.

When comments are posted, project participants are notified via email (left inset in Figure 2). Emails contain the body of the comment and clickable links that take users directly to the comment in a browser window. WebProtégé also provides integration with *Slack*[4] and makes it possible to associate a *Slack* Webhook[5] with a project so that comments are posted to a Slack channel for external notification. Slack users are able to click on links in their Slack clients to view the comment in WebProtégé (right inset in Figure 2).

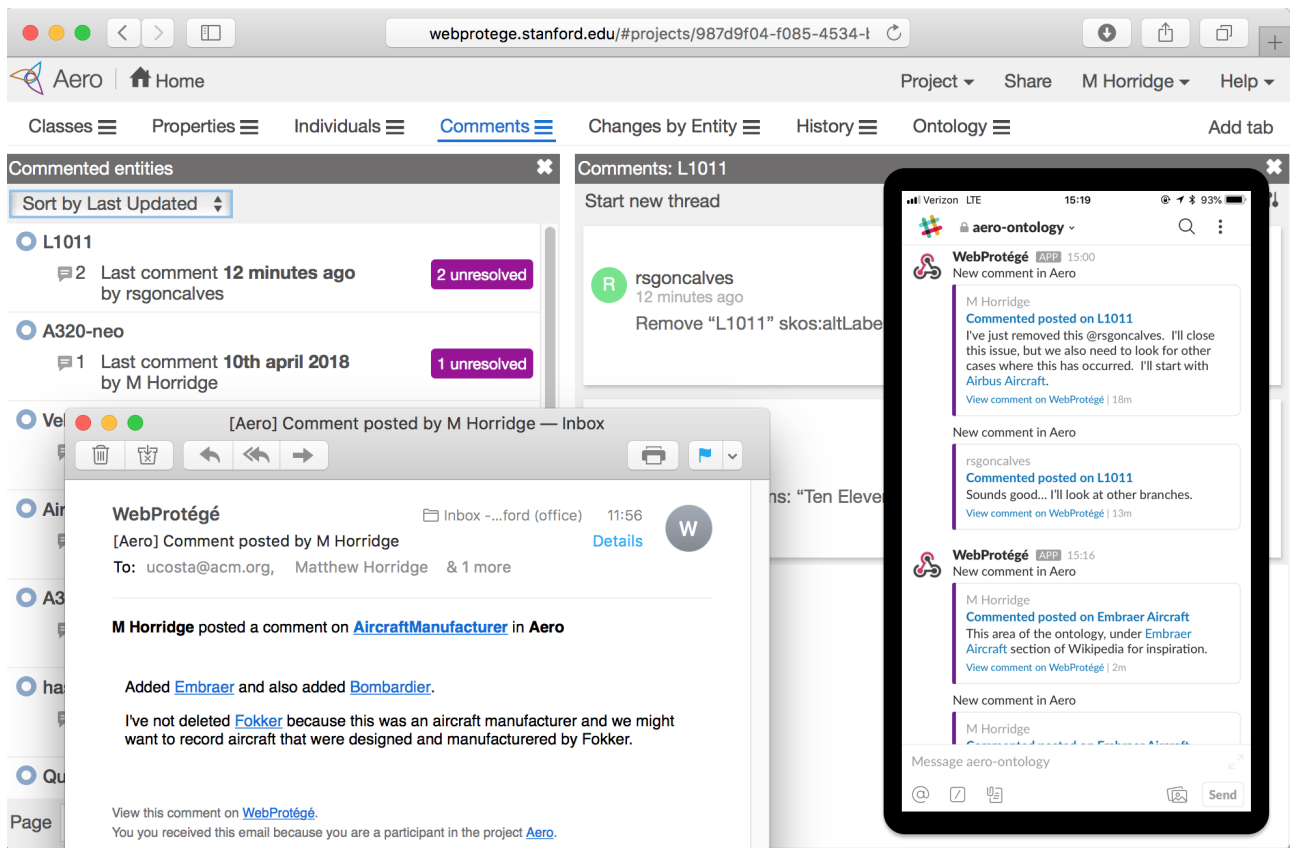

**Figure 2: Comments and Issues in WebProtégé. The main part of the figure shows the Comments tab. Comments can be sorted and viewed by entity. The lower left-hand side inset shows a notification email sent out to participants after a comment has been posted. The right-hand side inset shows integration with the chat app Slack.**

[4]https://slack.com
[5]https://api.slack.com/incoming-webhooks

In addition to comment threads, which can be used for adding review comments to a project, WebProtégé also allows entities to be "tagged" with tags that can be defined for a project. Figure 1 shows various tags in the class hierarchy that have been added as part of the review process. Tags and comments usually go hand-in-hand. Entities with specific tags can also be searched for in the search dialog. Entities can be automatically tagged according to rules, which users build using the same criteria that are available for searching entities (presented in Section 2.4). For example, entities that do not have a value for the rdfs:label annotation property could be automatically tagged with a tag Missing Label.

## 2.3 Change Tracking and Project History

One of the key features of WebProtégé is that all ontology changes are tracked and grouped into revisions. Revisions are based on atomic user interface operations, and it is possible to revert a single revision. Figure 3 shows an example of the history of a project. Changes can be viewed/filtered by entity, so that the entire history for an entity description (both logical and non-logical) can be examined. Labels for revisions are automatically generated based on the changes that occurred. In the full-blown OWL 2 editing interface it is possible to group a set of changes into a commit, which can then have a commit message assigned to it. WebProtégé allows downloading the set of ontologies in a given revision.

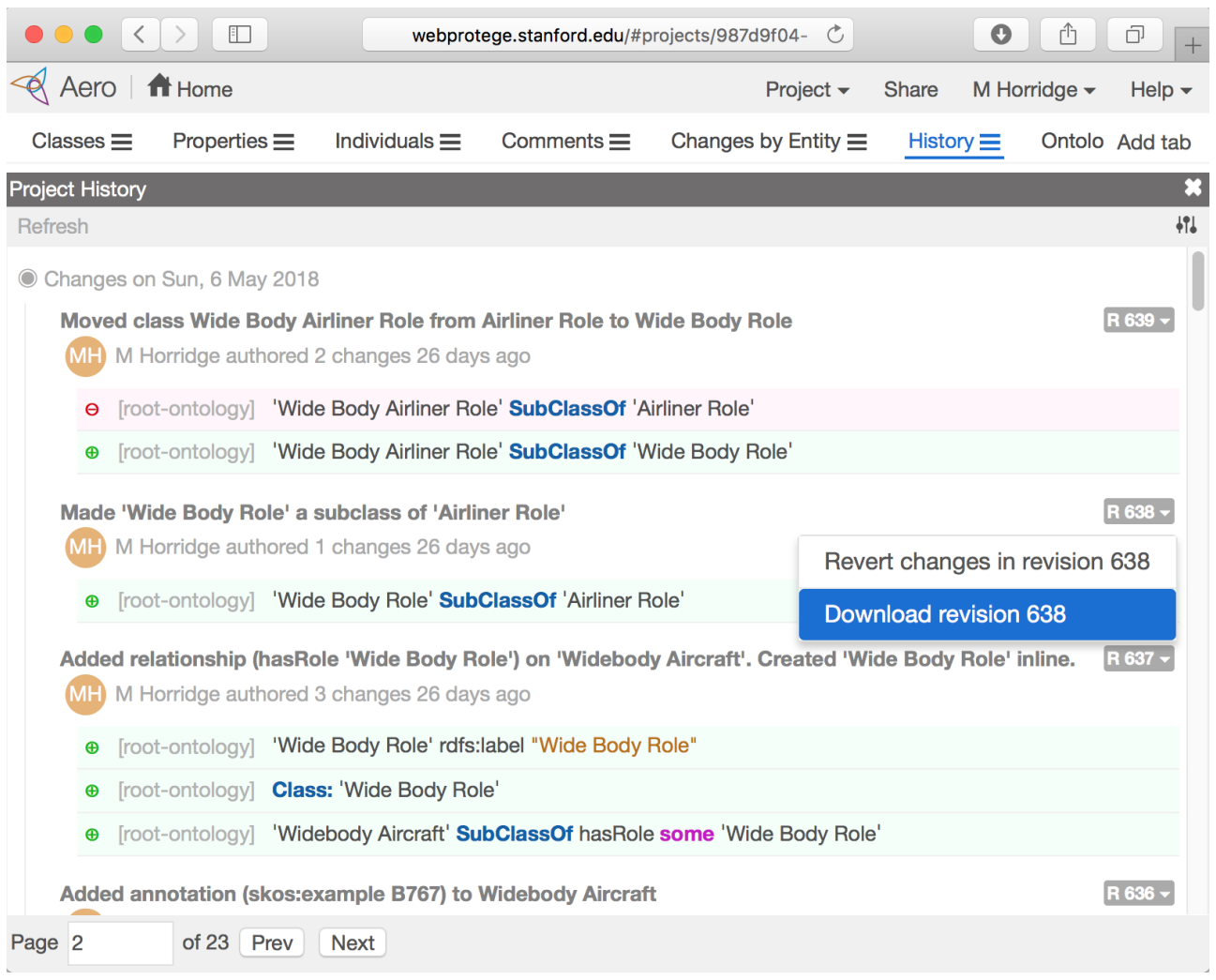

**Figure 3: Change Tracking in WebProtégé. All ontology changes are tracked and grouped into revisions that can be reverted and downloaded. Revisions are labelled with a description, author and timestamp.**

## 2.4 Querying

WebProtégé provides a simple user interface for performing complex ontology queries, shown in Figure 4. The kinds of queries supported essentially cover common search criteria and the basic expressivity of the SPARQL[6] query language.

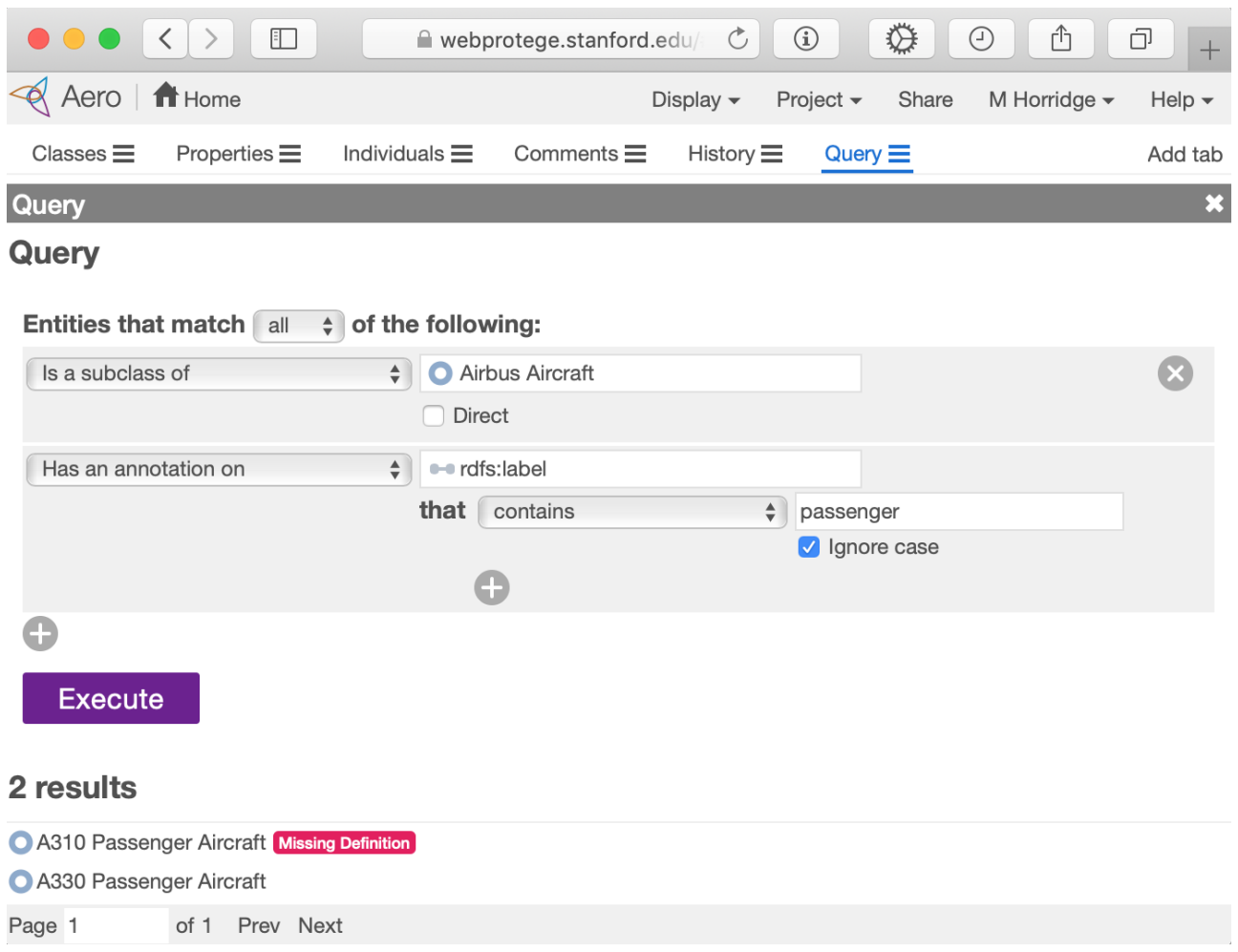

**Figure 4: Querying in WebProtégé. Users can build arbitrarily complex queries with commonly used search criteria. The figure exemplifies a query for entities that match *all* criteria: that entities are subclasses of Airbus Aircraft, and that entities have a value for the rdfs:label annotation property containing the string "passenger", regardless of text case.**

## 2.5 Graph Visualization

WebProtégé 4.0 features a brand new graph visualization, which we show in Figure 5. This visualization allows users to see all the relations that a selected entity takes part in. The unlabelled relations in the visualization represent rdfs:subClassOf relations, while labelled relations represent user-created properties. The unlabelled relations with a dashed line represent rdf:type relations. Using the graph visualization, users can isolate the relations (or "paths") in the graph between the selected entity and any entity that a user hovers over in the graph. Nodes in the graph can also be hidden from the visualization. WebProtégé allows users to download vector graphics files for the visualizations that it generates.

## 2.6 Integration with Third Party Applications

WebProtégé does not have a traditional "drop-in" plug-in mechanism like the desktop version of Protégé does. We intend to facilitate integration with third-party applications via Webhooks and a Web-API. As of now, third-party applications can be notified of the changes to a project by adding an event Webhook to a project. JSON messages are posted to the URL associated with the Webhook when changes in the project occur. We expect that this functionality will be used for loose coupling with third-party applications in the same way that GitHub and Slack can integrate with other applications.

[6]https://www.w3.org/TR/rdf-sparql-query

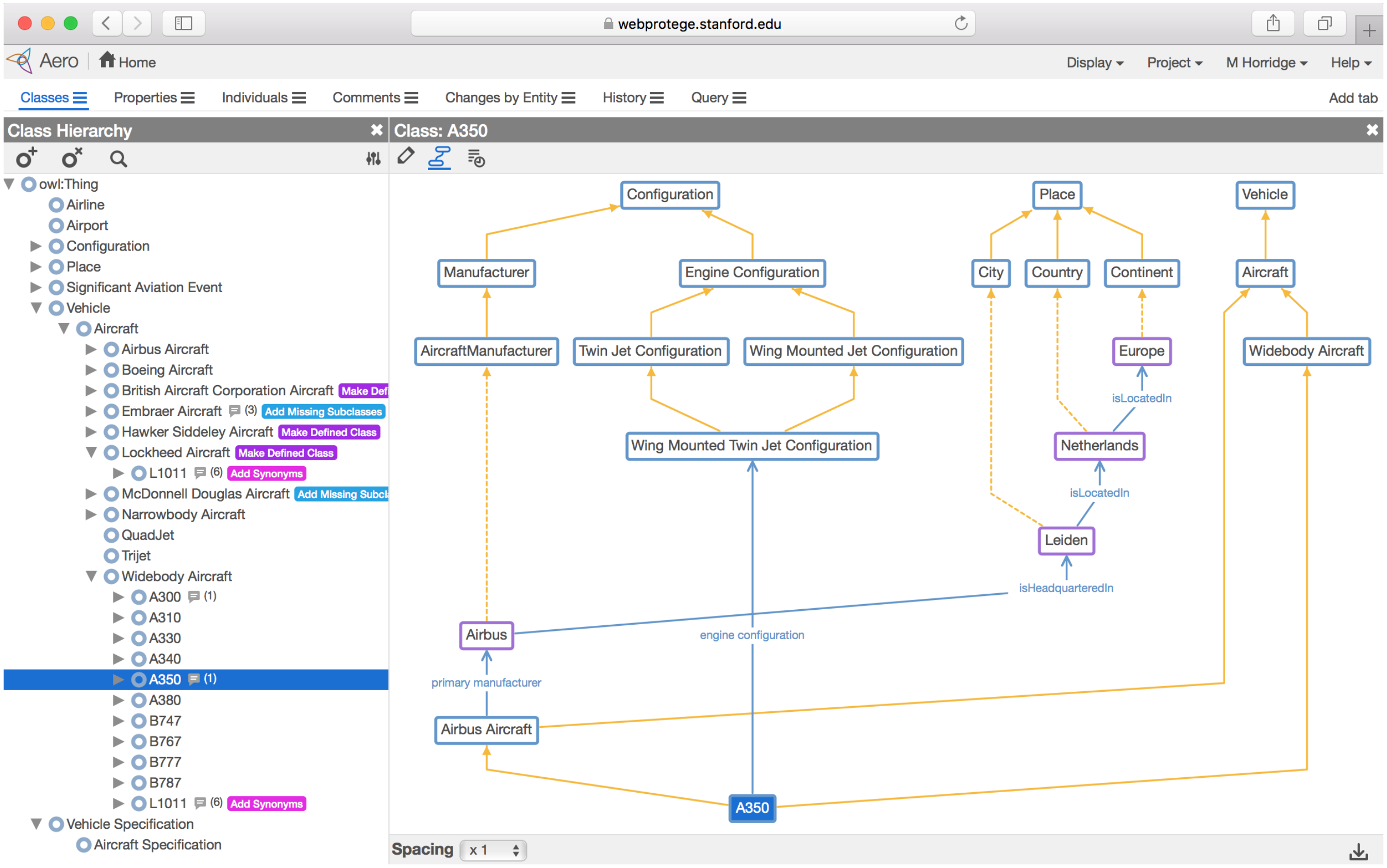

**Figure 5: WebProtégé graph visualization. This visualization is focused on the class A350, and shows rdfs:subClassOf relations in yellow, rdf:type relations in dashed yellow, and custom relations in blue, between A350 and other classes such as Configuration.**

## 3 SUMMARY

WebProtégé has been widely used since its inception. Important projects such as the 11th revision of the International Classification of Diseases (ICD-11) by the World Health Organization (WHO) depend on our software. The development environment used to edit ICD-11 is based on WebProtégé [4, 6]. We use WebProtégé to develop ontologies used in enterprise knowledge graphs. Some features such as search were inspired by these experiences.

The WebProtégé project would not be as successful as it is today without the engagement and the support of our massive user community. Our users help each other in the support mailing lists, and many of them interact closely with us on GitHub—by posting feature requests, issues, and prospective bug fixes or enhancements directly in our code repositories. We have also been fortunate to receive in-person feedback from our users as they learn how to develop ontologies using WebProtégé in our courses at Stanford University. Such feedback has proven to be invaluable in driving subsequent versions of the software.

In this paper, we have presented the major new features of the WebProtégé 4.0 ontology editor, developed at the Stanford Center for Biomedical Informatics Research. WebProtégé is under active development, and we seek feedback from the knowledge graph community so that we may continue to enhance WebProtégé to meet the needs of this community.

## ACKNOWLEDGEMENTS

This work was funded by Grant GM121724 from the National Institute of General Medical Sciences at the United States National Institutes of Health.